\documentclass[a4paper,11pt]{article}
\pdfoutput=1 % if your are submitting a pdflatex (i.e. if you have
\usepackage{jheppub} % for details on the use of the package, please
\usepackage[T1]{fontenc} % if needed
\usepackage{comment}

\usepackage[textsize=footnotesize,textwidth=2.cm]{todonotes}

\def\p{\partial}

\newcommand{\ba}{\begin{array}}
\newcommand{\ea}{\end{array}}
\newcommand{\bi}{\begin{itemize}}
\newcommand{\ei}{\end{itemize}}
\newcommand{\bea}{\begin{eqnarray}}
\newcommand{\eea}{\end{eqnarray}}
\newcommand{\be}{\begin{equation}}
\newcommand{\ee}{\end{equation}}

\title{\boldmath Correlation Functions of Warped CFT}

\author[a]{Wei Song}
\author{and}
\author[a]{Jianfei Xu}

\affiliation[a]{Yau Mathematical Sciences Center,Tsinghua University, Beijing, 100084, China}

\emailAdd{wsong@math.tsinghua.edu.cn}
\emailAdd{jfxu@math.tsinghua.edu.cn}

\abstract{Warped conformal field theory (WCFT) is a two dimensional quantum field theory whose local symmetry algebra consists of a Virasoro algebra and a $U(1)$ Kac-Moody algebra. In this paper, we study correlation functions for primary operators in WCFT. Similar to conformal symmetry, warped conformal symmetry is very constraining. The form of the two and three point functions are determined by the global warped conformal symmetry while the four point functions can be determined up to an arbitrary function of the cross ratio. The warped conformal bootstrap equation are constructed by formulating the notion of crossing symmetry. In the large central charge limit, four point functions can be decomposed into global warped conformal blocks,  which can be solved exactly.
Furthermore, we revisit the scattering problem in warped AdS spacetime (WAdS), and give a prescription on how to match the bulk result to a WCFT retarded Green's function. Our result is consistent with the conjectured holographic dualities between WCFT and WAdS.}

\begin{document}
\maketitle
\flushbottom
\section{Introduction}
Symmetry plays an essential role in quantum field theories. In conformal field theories, conformal symmetry is so constraining that it is possible to solve the theory via bootstrapping~\cite{Ferrara:1973yt, Polyakov:1974gs, Rattazzi:2008pe, Poland:2010wg, ElShowk:2012ht, El-Showk:2014dwa}. It is interesting to ask if these ideas work for symmetries other than conformal symmetry.
Recently a bootstrap for BMS symmetry (or Galileo conformal symmetry) was initiated~\cite{Bagchi:2016geg, Bagchi:2017cpu}. In this paper, we will focus on another  infinitely dimensional symmetry named the warped conformal symmetry, with a Warped Conformal Algebra (WCA) containing one Virasoro algebra plus one $U(1)$ Kac-Moody algebra.

From a purely field theoretical perspective, it was pointed out in~\cite{Hofman:2011zj} that a two dimensional quantum field theory with two global translational symmetries and a chiral global scaling symmetry have an extended local algebra. In contrast to the story with Lorentian symmetry and scaling symmetry~\cite{Pol}, there are two minimal options for this algebra. One is two copies of the Virasoro algebra and the other is a WCA. A field theory with the later choice is called the Warped Conformal Field Theory (WCFT)~\cite{Detournay:2012pc}. See~\cite{Detournay:2012pc} for a discussion about the representations and a warped Cardy formula, ~\cite{Compere:2013aya} for a bosonic example of WCFT, ~\cite{Hofman:2014loa, Castro:2015uaa} for Fermionic models,  and~\cite{Castro:2015csg, Song:2016gtd} for Entanglement entropy.

In this paper, we discuss the correlation functions of primary operators in WCFT. The form of the two and three point functions are determined by the global warped conformal symmetry while the four point functions can be determined up to an arbitrary function of the cross ratio. We will construct the warped conformal bootstrap equation by formulating the notion of crossing symmetry as in~\cite{Rattazzi:2008pe}. In the large central charge~\cite{Fitzpatrick:2014vua, Fitzpatrick:2015zha} and large Kac-Moody level limit, the four point functions can be decomposed into global warped conformal blocks, which can be calculated by solving the Casimir equations~\cite{Dolan:2000ut, Dolan:2003hv, SimmonsDuffin:2012uy}.
As a consistency check, these results can be applied to twist operators. Using Ward identities, the conformal dimension and $U(1)$ charge of twist fields are calculated,  from which the R$\acute{\mathrm{e}}$nyi entropy for an arbitrary single interval is obtained and confirms the results of~\cite{Castro:2015csg, Song:2016gtd}.

From a holographic perspective, WCFT originated from the search for holographic dual to a class of geometries with $SL(2, R)\times U(1)$ isometry, including warped AdS$_3$ ( WAdS)~\cite{Anninos:2008fx} and the near horizon of extremal Kerr( NHEK) \cite{Bardeen:1999px, Guica:2008mu}. Under Dirichlet-Neumann boundary conditions~\cite{Compere:2009zj}, holographic dual for asymptotically WAdS spacetime will fulfils the warped conformal symmetry. Such boundary conditions can also be imposed to AdS$_3$~\cite{Compere:2013bya}, suggesting the possibility of WCFT as an alternative holographic dual to AdS$_3$, i.e., AdS$_3$/WCFT. The evidence of the WAdS/WCFT conjecture comes from the matching of the WAdS black hole entropy and the microscopic entropy given by a Cardy-like formula~\cite{Detournay:2012pc}. A bulk calculation of the entanglement entropy in the contexts of AdS$_3$/WCFT and WAdS/WCFT has been worked out in~\cite{Song:2016gtd}, using a generalization of the Rindler method~\cite{Casini:2011kv}. However, a matching between a bulk scattering problem and a retarded Green's function, analogous to~\cite{Bredberg:2009pv}, is still missing.

In this paper, we provide a prescription on how the dictionary of WAdS/WCFT should work for correlation functions. The main point is that the momentum conjugate to the $U(1)$ isometry, which is promoted to a current algebra at the boundary, should be understood as a charge. An operator with fixed charge correspond to a field with fixed momentum. With this prescription, the retarded Green's function with fixed momentum calculated in WAdS matches the thermal two point function of an operator with fixed charge in WCFT.

 %a proper bulk coordinate transformation which maps the locally AdS$_3$ or WAdS to the analogy of a hyperbolic black hole in three dimensions. And the corresponding boundary coordinate transformation is a general version of a Rindler transformation mentioned in~\cite{Castro:2015csg}. The hyperbolic black hole entropy matches the thermal entropy for the Rindler space which gives a natural bulk interpretation of the entanglement entropy for WCFT.

The layout of the paper is the following. In section \ref{sec2}, correlation functions will be discussed. The warped conformal bootstrap equation based on the crossing symmetry will be given. In section \ref{sec3}, thermal correlators will be discussed. In section \ref{sec4}, we revisit the calculation of the R$\acute{\mathrm{e}}$nyi entropy for a single interval in WCFT. In section \ref{sec5}, we reinterpret scattering problem in WAdS as retarded Green's functions in WCFT.

\section{Correlation Functions from Warped CFT}\label{sec2}
\subsection{The Warped Conformal Symmetry}
A warped conformal field theory is characterized by the warped conformal symmetry. The global symmetries are $SL(2,R)\times U(1)$,  while the local symmetry algebra is a Virasoro algebra plus a $U(1)$ Kac-Moody algebra.
In position space, a general warped conformal symmetry transformation can be written as
\begin{equation}\label{wtr}
x'=f(x),~~~~y'=y+g(x)\,,
\end{equation}
where $f(x)$ and $g(x)$ are two arbitrary functions.

Consider a WCFT on a plane. Infinitesimally, the warped conformal symmetry are generated by a set of vector fields,
\begin{equation}\label{wg}
L_n=-x^{n+1}\partial_x,~~~~P_n=ix^n\partial_y\,.
\end{equation}
These vector fields form a Virasoro-Kac-Moody algebra,
\begin{align}\label{wca}
[L_n, L_m]=&(n-m)L_{n+m}\,,\nonumber\\
[L_n, P_m]=&-mP_{n+m}\,,\nonumber\\
[P_n, P_m]=&0\,.
\end{align}
Global symmetry transformations are generated by $L_{\pm 1},\,L_0$ and $P_0$, which form a $SL(2,R)\otimes U(1)$ sub-algebra. In particular, scaling symmetry is generated by $L_0$, and translation symmetries are generated by $L_{-1}$ and $P_0$.

Denote $T(x)$ and $P(x)$ as the Noether currents associated with translations along $x$ and $y$ axis, respectively. There are infinitely many conserved charges~\cite{Hofman:2011zj,Detournay:2012pc},
\begin{equation}
\mathcal{L}_n=-\frac{i}{2\pi}\int\mathrm{d}xx^{n+1}T(x),~~~~\mathcal{P}_n=-\frac{1}{2\pi}\int\mathrm{d}xx^nP(x)\,.
\end{equation}
The commutation relations for the charges form a warped conformal algebra consists of one Virasoro algebra and a Kac-Moody algebra,
\begin{align}\label{wcack}
[\mathcal{L}_n, \mathcal{L}_m]=&(n-m)\mathcal{L}_{n+m}+\frac{c}{12}n(n^2-1)\delta_{n, -m}\,,\nonumber\\
[\mathcal{L}_n, \mathcal{P}_m]=&-m\mathcal{P}_{n+m}\,,\nonumber\\
[\mathcal{P}_n, \mathcal{P}_m]=&k\frac{n}{2}\delta_{n, -m}\,,
\end{align}
where $c$ is the central charge and $k$ is the Kac-Moody level. The finite transformation properties of the energy momentum tensor and Kac-Moody current are given by,
\begin{eqnarray}\label{Tp}
T'(x')&=&\left(\frac{\partial x}{\partial x'}\right)^2\left(T(x)-\frac{c}{12}\{x', x\}\right)+\frac{\partial x}{\partial x'}\frac{\partial y}{\partial x'}P(x)-\frac{k}{4}\left(\frac{\partial y}{\partial x'}\right)^2\\\label{Pp}
P'(x')&=&\frac{\partial x}{\partial x'}\left(P(x)+\frac{k}{2}\frac{\partial y'}{\partial x}\right)\,,
\end{eqnarray}
where
\begin{equation}
\{x', x\}=\frac{\frac{\partial^3x'}{\partial x^3}}{\frac{\partial x'}{\partial x}}-\frac{3}{2}\left(\frac{\frac{\partial^2x'}{\partial x^2}}{\frac{\partial x'}{\partial x}}\right)^2\,.
\end{equation}
One can also construct the spectral flow invariant Virasoro generators,
\bea
{\mathcal{L}}^{inv}_{n}&=&\mathcal{L}_n-{1\over k} \Big(\sum_{m\le-1}\mathcal{P}_{m}\mathcal{P}_{n-m}+\sum_{m\ge0}\mathcal{P}_{n-m}\mathcal{P}_m\Big)\,.
\eea
One can check that ${\mathcal{L}}^{inv}_{n}$ commutate with the Kac-Moody generators, and
form a Virasoro algebra with conformal weight $c-1$.

\subsection{State Operator Correspondence}
Following \cite{Hofman:2011zj}, we start with a field theory with global scaling symmetry $x\rightarrow \lambda x$ and translational symmetry
$x\rightarrow x+a,\,y\rightarrow y+b$,
and assume the existence of a complete basis of  local operators
$\Phi_i(x,y)$ transform as
\bea
\Phi'_i(\lambda x+a,y+b)=\lambda^{-\Delta_i}\Phi_i(x,y)\,,
\eea
where $\Delta_i$ is the scaling dimension of the operator $\Phi_i(x, y)$. Or equivalently, we can write the infinitesimal transformations as
\bea \label{globalrule}\,[\mathcal{L}_0, \Phi_i(x,y)]&=&(x\p_x+\Delta_i )\Phi_i (x,y),\\
\,[\mathcal{L}_{-1}, \Phi_i(x,y)]&=&\p_x\Phi_i(x,y),
\\
\,\label{P0p}[\mathcal{P}_0, \Phi_i(x,y)]&=&i \p_y \Phi_i(x,y)\,.
 \eea
With this setup, \cite{Hofman:2011zj} argues that there are two minimal choices for the local symmetries, one choice leads to two Virasoro algebras and the other leads to a warped conformal algebra (\ref{wcack}). Specializing to WCFT, since $\mathcal{P}_0$ commute with all the other generators, we can furthermore require the basis to have definite $\mathcal{P}_0$ charge, namely,
\be \,[\mathcal{P}_0, \Phi_i(x,y)]=-Q_i\Phi_i(x,y)\label{P0QPhi}\,.\ee
In fact, due to equations (\ref{P0p}) and (\ref{P0QPhi}), the $y$ dependence in the local operator $\Phi_i(x,y)$ is determined, i.e., $\Phi_i(x,y)=e^{iQ_iy}\Phi_i(x)$, where $\Phi_i(x)$ is only a function of $x$. Indeed, such property determines the $y$ part of all correlation functions as we will see in the following.

Similarly to CFT story, we define primary operators by their transformation rules,
\bea \label{poft}
\mathcal{O}'(x', y')=\left(\frac{\partial x'}{\partial x}\right)^{-\Delta}\mathcal{O}(x, y)\,.
 \eea
where $\Delta$ is the scaling dimension of the primary operator $\mathcal{O}(x, y)$.
The infinitesimal version is,
\begin{align}
[\mathcal{L}_n, \mathcal{O}(x, y)]=&[x^{n+1}\partial_x+(n+1)x^n\Delta]\mathcal{O}(x, y)\,,\label{Ln}\\
[\mathcal{P}_n, \mathcal{O}(x, y)]=&ix^n\p_y\mathcal{O}(x, y)\,\label{Pn}\\
=&-x^n Q \mathcal{O}(x, y)\,,\label{P0Q}
\end{align}
where the last line is due to the fact that we are choosing a basis with definite $\mathcal{P}_0$ charge $Q$ (\ref{P0QPhi}).
%A descendant operator is defined as
%$\mathcal{O}^{\{\vec{N}, \vec{M}\}}= \mathcal{L}_{-1}^{N_1} \mathcal{L}_{-2}^{N_2}...\mathcal{P}_{-1}^{{M}_1}\mathcal{P}_{-2}^{{M}_2}...\mathcal{O}\,,$
%where $\vec{N}={N_1,\, N_2,\cdots}$, and $\vec{M}={M_1,\, M_2,\cdots}$.
%Using the commutation relations,
%the $U(1)$ charge of $\mathcal{O}^{\{\vec{N}, \vec{M}\}}$ is still $Q$ and the conformal weight becomes
%\begin{equation}
%h=\Delta+\sum_{n>0} nN_n+\sum_{m>0} {m} {M}_m\,.
%\end{equation}
%Note that throughout this paper, we use $h$ to denote the scaling dimension of a generic operator, and $\Delta$ for a primary operator. %Written in terms of the spectral flow invariant operators,
%we have,
%\bea  [\mathcal{L}^{inv}_n, \mathcal{O}(x, y)]=&[x^{n+1}\partial_x+(n+1)x^n\Delta^{inv}]\mathcal{O}(x, y)\,,\label{Lninv}
%\eea
%where $\Delta^{inv}=\Delta-\frac{Q^2}{k}$. Alternatively, we can choose another basis,
%\bea\label{DOinv}
%\mathcal{O}^{\{\vec{N}, \vec{M}\}}= \left(\mathcal{L}_{-1}^{inv}\right)^{N_1} \left(\mathcal{L}_{-2}^{inv}\right)^{N_2}...\mathcal{P}_{-1}^{{M}_1}\mathcal{P}_{-2}^{{M}_2}...\mathcal{O}\,.
%\eea

To discuss state operator correspondence, we consider a complex mapping between plane and cylinder
\footnote{More discussions about the representation and  the complex mapping can be found in section 2 of both \cite{Detournay:2012pc} and \cite{Castro:2015uaa}}\bea
x=e^{-i(t-\phi)}=e^{t_E+i\phi}\,,\quad y=\phi+\gamma (t-\phi)\,,
\eea
where $t$ is interpreted as Lorentzian time, $\phi\sim \phi+2\pi$ as a spatial circle, $t_E\rightarrow -it$ as a Euclidean time. $\gamma$ is a spectral flow parameter, and leads to different interpretations of the $U(1)$ direction.  The meaning of this complex mapping and analytic continuation is that the Lorentzian cylinder parameterized by $t,\phi$ is capped off at $t=0$ by a Euclidean disk. A radial quantization can be performed on the complex plane, and the states are glued it to states on the cylinder.
Having an initial state at very early Euclidean time corresponding to insert an operator at $x=0$, and vice versa. Using translational symmetry, we can further put the operator at $y=0$.
In particular, a primary operator with weight $\Delta$ and charge $Q$ at $x=0$ defines a state,
\be \mathcal{O}(0,0)\sim |\Delta, Q\rangle\,.\ee
Using (\ref{Ln}), (\ref{Pn}), we have,
\bea
\mathcal{L}_0|\Delta, Q\rangle&=&\Delta|\Delta, Q\rangle,\quad \mathcal{P}_0|\Delta, Q\rangle=-Q|\Delta, Q\rangle,\\
\label{pr}
\mathcal{L}_n|\Delta, Q\rangle&=&0,~~~~\mathcal{P}_n|\Delta, Q\rangle=0,~~~~\forall n>0\,,
\eea
which is just the definition of primary states discussed in~\cite{Detournay:2012pc,Compere:2013bya}, but now rewritten on the plane. In particular, the unit operator corresponds to the $SL(2,R)\times U(1)$ invariant vacuum. Similarly, the descendent operators also correspond to descendent states. In general, vacuum charges on the cylinder will be non-zero due to the central charge and the spectral flow.

So far we have shown that given a definition of a complete basis of operators, we can find a complete basis of states. As a consistency check, we can reverse the process. A complete basis of states defines a set of operators at the origin.
Primary states corresponds to an operator at origin with,
\begin{equation}\label{prim}
[\mathcal{L}_0, \mathcal{O}(0,0)]=\Delta\mathcal{O},~[\mathcal{P}_0, \mathcal{O}(0,0)]=-Q\mathcal{O}(0,0),~[\mathcal{L}_n, \mathcal{O}(0,0)]=0,~[\mathcal{P}_n, \mathcal{O}(0,0)]=0,~~~~\forall n>0\,.
\end{equation}
If we define local operator by
\be \mathcal{O}(x, y)=U^{-1}\mathcal{O}(0, 0)U,~~~~\text{where}~~U=e^{-x\mathcal{L}_{-1}-iy\mathcal{P}_0}\,.\ee
Using the commutation relations (see appendix \ref{appA}), one can show that this operator indeed satisfy the transformation rule (\ref{Ln})-(\ref{P0Q}).

It is easy to check that the primary state $|\Delta, Q\rangle$ is also a primary state under $\mathcal{L}^{inv}_n$ and $\mathcal{P}_n$, \bea
\mathcal{L}^{inv}_0|\Delta, Q\rangle&=&\Delta^{inv}|\Delta, Q\rangle,\quad \mathcal{P}_0|\Delta, Q\rangle=-Q|\Delta, Q\rangle\,,\\
\label{pr}
\mathcal{L}^{inv}_n|\Delta, Q\rangle&=&0,~~~~\mathcal{P}_n|\Delta, Q\rangle=0,~~~~\forall n>0\,.\\
\eea
where the conformal weight is
\be \Delta^{inv}=\Delta-{Q^2\over k}\,. \ee
Since the spectral flow invariant Virasoro $\mathcal{L}^{inv}_n$ and $\mathcal{P}_n$ commute, sometimes it is more convenient to label states using the eigenvalues $\Delta^{inv}$ and $Q$, and define the descendants using the basis,
\bea
|\mathcal{O}^{\{\vec{N}, \vec{M}\}}\rangle \equiv \left(\mathcal{L}_{-1}^{inv}\right)^{N_1} \left(\mathcal{L}_{-2}^{inv}\right)^{N_2}...\mathcal{P}_{-1}^{{M}_1}\mathcal{P}_{-2}^{{M}_2}...|\Delta, Q\rangle\,.\label{ONM}
\eea
The spectral invariant conformal weight and charge for such a descendant state are,
\bea
\mathcal{L}_0^{inv}|\mathcal{O}^{\{\vec{N}, \vec{M}\}}\rangle &=& \left(\Delta^{inv}+\sum_{n>0} n N_n\right)|\mathcal{O}^{\{\vec{N}, \vec{M}\}}\rangle, \\
\mathcal{P}_0|\mathcal{O}^{\{\vec{N}, \vec{M}\}}\rangle &=& -Q |\mathcal{O}^{\{\vec{N}, \vec{M}\}}\rangle\,.
\eea

\subsection{Two point functions}
The two point functions can be fixed by the global symmetries that generated by $\mathcal{L}_{0, \pm1}$ and $\mathcal{P}_0$.
Results for two point functions were first derived in a slightly different way in \cite{Castro:2015csg}. Here in this subsection, we rederive it and meanwhile lay out a basis for three and four point functions.
Consider two primary operators $\mathcal{O}_1(x_1, y_1)$ and $\mathcal{O}_2(x_2, y_2)$ with conformal weights and charges $(\Delta_1, Q_1)$ and $(\Delta_2, Q_2)$ respectively. The two point function is defined as,
\begin{equation}
G^{(2)}(x_1, x_2, y_1, y_2)=\langle0|T\mathcal{O}_1(x_1, y_1)\mathcal{O}_2(x_2, y_2)|0\rangle\,,
\end{equation}
where $T$ stands for time ordering. Throughout this paper, we will always assume $x_1>x_2$. Since the vacuum state $|0\rangle$ is $SL(2,R)\times U(1)$ invariant, the two point function is invariant under the action of $\mathcal{L}_{0, \pm1}$ and $\mathcal{P}_0$.
The action of the $SL(2, R)$ part, i.e., $\mathcal{L}_{0, \pm1}$ fix the $x$ dependence of the two point function,
\begin{equation}
G^{(2)}(x_1, x_2, y_1, y_2)=f(y_1, y_2)\delta_{\Delta_1, \Delta_2}\frac{1}{(x_1-x_2)^{2\Delta_1}}\,,
\end{equation}
where $f(y_1, y_2)$ is an arbitrary function of $y_1$ and $y_2$. The action of $\mathcal{P}_0$, i.e., (\ref{P0Q}) leads to the charge conservation condition,
\begin{equation}
\langle0|[\mathcal{P}_0, G^{(2)}]|0\rangle=0\Rightarrow Q_1+Q_2=0\,.
\end{equation}
and the right hand sides of (\ref{Pn}) and (\ref{P0Q}) implies
\begin{align}
&(\partial_y-iQ)\mathcal{O}(x, y)=0\nonumber\\
\Rightarrow&(\partial_{y_j}-iQ_j)G^{(2)}(x_1, x_2, y_1, y_2)=0,~~~~j=1,2\nonumber\\
\Rightarrow&G^{(2)}(x_1, x_2, y_1, y_2)=C^{(2)}\delta_{\Delta_1, \Delta_2}e^{iQ_1y_1+iQ_2y_2}\frac{1}{(x_1-x_2)^{2\Delta_1}}\,,
\end{align}
where $C^{(2)}$ is an arbitrary constant.  To wrap up, the normalized two point function of the WCA reads,
\begin{equation}\label{2pf0}
G^{(2)}(x_1, x_2, y_1, y_2)=\delta_{\Delta_1, \Delta_2}\delta_{Q_1, -Q_2}e^{iQ_1(y_1-y_2)}\frac{1}{(x_1-x_2)^{2\Delta_1}}\,.
\end{equation}
The two point correlator is non-zero only when the scaling dimensions are identical and the charges add up to zero.

According to (\ref{poft}), under finite transformations (\ref{wtr}), the two point function for the primary operator transforms as,
\be\label{ftG}
G'^{(2)}(x'_1, x'_2, y'_1, y'_2)=\left(\frac{\partial x'_1}{\partial x_1}\right)^{-\Delta_1}\left(\frac{\partial x'_2}{\partial x_2}\right)^{-\Delta_2}G^{(2)}(x_1, x_2, y_1, y_2)\,.
\ee
In particular, we can perform a spectral flow $x'= x$, $y'= y-qx$. Then the vacuum after the spectral flow is still annihilated by the original $SL(2,R)\times U(1)$ generators, which can be rewritten in the new coordinates,
\be\label{lpq}
L_n=-x'^{n+1}(\p_x'-q\p_y'),~~~~P_n=ix'^n\p_y'\,.
\ee
The two point function is then
\be
G'^{(2)}(x'_1, x'_2, y'_1, y'_2)=\delta_{\Delta_1, \Delta_2}\delta_{Q_1, -Q_2}e^{iQ_1(y'_1-y'_2-q(x'_1-x'_2))}\frac{1}{(x'_1-x'_2)^{2\Delta_1}}\,.
\ee
This agrees with the Eq.(3.46) of \cite{Castro:2015csg}, with a choice of $q$ which will be determined later.
In fact, the results in $(x,y)$ system should be understood as the spectral flow invariant expressions.
Results in other frame can be obtained by a spectral flow.

\subsection{Three point functions}
By the same token of symmetry analysis, one can construct three point function for primary operators $\mathcal{O}_i(\Delta=\Delta_i, Q=Q_i), i=1, 2, 3$. The three point function is defined as,
\begin{equation}
G^{(3)}(x_1, x_2, x_3, y_1, y_2, y_3)=\langle0|T\mathcal{O}_1(x_1, y_1)\mathcal{O}_2(x_2, y_2)\mathcal{O}_3(x_3, y_3)|0\rangle\,.
\end{equation}
The action of the $SL(2, R)$ global symmetry requires that the $x$ dependence of the three point function should be,
\begin{equation}
G^{(3)}=f(y_1, y_2, y_3)\frac{1}{x_{12}^{\Delta_1+\Delta_2-\Delta_3}x_{31}^{\Delta_3+\Delta_1-\Delta_2}x_{23}^{\Delta_2+\Delta_3-\Delta_1}}\,,
\end{equation}
where $x_{ij}=x_i-x_j$, $f(y_1, y_2, y_3)$ is an arbitrary function of $y_1$, $y_2$, and $y_3$. The action of $\mathcal{P}_0$ leads to the charge conservation condition for the three point function, $Q_1+Q_2+Q_3=0$.
More symmetrically, the three point function of the WCA can be written as,
\begin{align}\label{3pf0}
G^{(3)}&=C^{(3)}\delta_{Q_1+Q_2+Q_3, 0}\frac{e^{\frac{i}{3}(Q_1-Q_2)y_{12}}}{x_{12}^{\Delta_1+\Delta_2-\Delta_3}}\frac{e^{\frac{i}{3}(Q_3-Q_1)y_{31}}}{x_{31}^{\Delta_3+\Delta_1-\Delta_2}}\frac{e^{\frac{i}{3}(Q_2-Q_3)y_{23}}}{x_{23}^{\Delta_2+\Delta_3-\Delta_1}}\nonumber\\
&=C^{(3)}\delta_{Q_1+Q_2+Q_3, 0}\frac{x_{12}^{\frac{2Q_1Q_2}{k}}e^{\frac{i}{3}(Q_1-Q_2)y_{12}}}{x_{12}^{\Delta_1^{inv}+\Delta_2^{inv}-\Delta_3^{inv}}}\frac{x_{31}^{\frac{2Q_3Q_1}{k}}e^{\frac{i}{3}(Q_3-Q_1)y_{31}}}{x_{31}^{\Delta_3^{inv}+\Delta_1^{inv}-\Delta_2^{inv}}}\frac{x_{23}^{\frac{2Q_2Q_3}{k}}e^{\frac{i}{3}(Q_2-Q_3)y_{23}}}{x_{23}^{\Delta_2^{inv}+\Delta_3^{inv}-\Delta_1^{inv}}}\,,
\end{align}
where $y_{ij}=y_i=y_j$,  and $C^{(3)}$ is an arbitrary constant. Once the two point function is normalized as in Eq. (\ref{2pf0}), the constant $C^{(3)}$ equals to the OPE coefficient $C_{\mathcal{O}_1\mathcal{O}_2\mathcal{O}_3}$.

We can also consider the three point function of two operators and a Virasoro descendent,
\bea
%&&\langle \mathcal{O}_1(\infty,y_1) \mathcal{L}_n^{inv}\mathcal{O}_2(x,y_2) \mathcal{O}_3(0,y_3)\rangle \nonumber\\
&&\langle \Delta, Q | \mathcal{L}_n^{inv}\mathcal{O}_2(x,y_2) \mathcal{O}_3(0,y_3)\rangle \nonumber\\
&=&\langle \Delta, Q | \mathcal{L}_n-{1\over k}\Big(\mathcal{P}_0\mathcal{P}_n+\mathcal{P}_1\mathcal{P}_{n-1}+\cdots \mathcal{P}_{n-1}\mathcal{P}_1 +\mathcal{P}_n\mathcal{P}_0\Big)\mathcal{O}_2(x,y_2) \mathcal{O}_3(0,y_3)\rangle \nonumber\\
&=&\langle \Delta, Q | \Big(\Delta^{inv}_2(n+1)x^n+x^{n+1}\p_x-{2Q_2 Q_3\over k}x^n\Big)\mathcal{O}_2(x,y_2) \mathcal{O}_3(0,y_3)\rangle\nonumber\\
&=& x^{2Q_2 Q_3/k}[\langle \Delta, Q | \mathcal{L}_n\mathcal{O}_2(x,y_2) \mathcal{O}_3(0,y_3)\rangle]_{\Delta_i\rightarrow \Delta_i^{inv}}\,. \label{inv3pt}
\eea
One can show that descendants with multiple $\mathcal{L}_n^{inv}$ leads to a similar result with a universal pre-factor independent of conformal weights.
Since the spetral flow invariant Virasoro and Kac-Moody generators commute, the action of $\mathcal{L}_n^{inv}$ and $\mathcal{P}_m$ factors.

\subsection{Four point functions and warped conformal bootstrap}
Like in CFTs, the four point functions in warped conformal field theory are not completely determined by the global subgroup $\{\mathcal{L}_{0, \pm1}, \mathcal{P}_0\}$. They depend on arbitrary functions of the cross ratio $x$ given by,
\begin{equation}
x=\frac{x_{12}x_{34}}{x_{13}x_{24}}\,,
\end{equation}
This cross ratio is invariant under the global generators $\{L_{0, \pm1}, P_0\}$ defined in (\ref{wg}). The global symmetry requires that the four point function takes the form,
\begin{align}
\langle\prod_{j=1}^{4}\mathcal{O}_j(x_j, y_j)\rangle=\left(\prod_{j<k}e^{\frac{i}{4}(Q_j-Q_k)y_{jk}}\right)\left(\frac{x_{24}}{x_{14}}\right)^{\Delta_1-\Delta_2}\left(\frac{x_{14}}{x_{13}}\right)^{\Delta_3-\Delta_4}\frac{\mathcal{G}(x)}{x_{12}^{\Delta_1+\Delta_2}x_{34}^{\Delta_3+\Delta_4}}\,,
\end{align}
where $\Delta_i$ and $Q_i$ are conformal weights and charges for the primary operators $\mathcal{O}_i$, and $\mathcal{G}(x)$ is an undetermined function of $x$. The action of $\mathcal{P}_0$ on the four point function followed by (\ref{P0Q}) leads to the charge conservation condition, i.e., $\sum_{i=1}^4Q_i=0$.

A complete basis of states can be decomposed into primary states and their warped conformal descendants. By operator state correspondence, a complete basis of operators can also be decomposed similarly.
By inserting a complete basis, four point functions can be decomposed as a sum over conformal partial waves,
\bea
\langle\prod_{j=1}^{4}\mathcal{O}_j(x_j, y_j)\rangle=\sum_{{\mathcal O}}C_{12\mathcal{O}}C^{\mathcal{O}}_{~~34} W^{\Delta, Q}(x_i, y_i)\,.
\eea
Warped conformal partial waves are defined as,
\begin{equation}\label{Wp}
W^{\Delta, Q}(x_i, y_i)=\frac{1}{C_{12\mathcal{O}}C^{\mathcal{O}}_{~~34}}\langle\mathcal{O}_1(x_1, y_1)\mathcal{O}_2(x_2, y_2)\Big(\sum_{\vec{N},\vec{M}}{|\mathcal{O}^{\{\vec{N}, \vec{M}\}}\rangle \langle\mathcal{O}^{\{\vec{N}, \vec{M}\}}|\over \mathcal{N}_{\vec{N},\vec{M}}}\Big)\mathcal{O}_3(x_3, y_3)\mathcal{O}_4(x_4, y_4)\rangle\,,
\end{equation}
where $\mathcal{O}$ is a primary of dimension $\Delta$ and charge $Q$, $|\mathcal{O}^{\{\vec{N}, \vec{M}\}}\rangle$ stands for all Virasoro-Kac-Moody descendant states (\ref{ONM}) with normalization $\mathcal{N}_{\vec{N},\vec{M}}$, and $C_{12\mathcal{O}}$ and $C^{\mathcal{O}}_{~~34}$ are the OPE coefficients. Accordingly, the function $\mathcal{G}(x)$ can be decomposed into warped conformal blocks $A^{\Delta, Q}(x)$~\cite{Ferrara:1973yt, Polyakov:1974gs},
\begin{equation}\label{cbd}
\mathcal{G}(x)=\sum_{\mathcal{O}}C_{12\mathcal{O}}C^{\mathcal{O}}_{~~34}A^{\Delta, Q}(x)\,.
\end{equation}
where the warped conformal block is related to the warped conformal partial wave by,
\begin{equation}\label{pw}
W^{\Delta, Q}(x_i, y_i)\equiv\left(\prod_{j<k}e^{\frac{i}{4}(Q_j-Q_k)y_{jk}}\right)\left(\frac{x_{24}}{x_{14}}\right)^{\Delta_1-\Delta_2}\left(\frac{x_{14}}{x_{13}}\right)^{\Delta_3-\Delta_4}\frac{A^{\Delta, Q}(x)}{x_{12}^{\Delta_1+\Delta_2}x_{34}^{\Delta_3+\Delta_4}}\,.
\end{equation}
We recognize that the warped conformal block is just the Virasoro-Kac-Moody block discussed in \cite{Fitzpatrick:2015zha}.
Due to the global warped conformal symmetry, we can do a coordinate transformation to sit the four points at,
\begin{equation}
\{(x_1, y_1), (x_2, y_2), (x_3, y_3), (x_4, y_4)\}\to\{(\infty, 0), (1, y_2), (x, y), (0, y_4)\}\,.
\end{equation}
and define
\begin{equation}\label{G1234}
G^{21}_{34}(x)\equiv\lim_{x_1\to\infty, y_1\to0}e^{i(Q_2y_2+Q_3y+Q_4y_4)}x_1^{2\Delta_1}\langle\mathcal{O}_1(x_1, y_1)\mathcal{O}_2(1, y_2)\mathcal{O}_3(x, y)\mathcal{O}_4(0, y_4)\rangle\,.
\end{equation}
Note that $G^{21}_{34}(x)$ is a function of the cross ratio $x$ which can be directly decomposed into conformal blocks as (\ref{cbd}). The four point function is invariant under the ordering of operators inside. Similarly, we can also define,
\begin{equation}\label{G1432}
G^{41}_{32}(x)\equiv\lim_{x_1\to\infty, y_1\to0}e^{i(Q_2y_4+Q_3y+Q_4y_2)}x_1^{2\Delta_1}\langle\mathcal{O}_1(x_1, y_1)\mathcal{O}_4(1, y_2)\mathcal{O}_3(x, y)\mathcal{O}_2(0, y_4)\rangle\,.
\end{equation}
The crossing symmetry from exchanging points 2 and 4 is given by the following equation,
\begin{equation}\label{crosssym}
G^{21}_{34}(x)=G^{41}_{32}(1-x)\,.
\end{equation}
Using the OPE between $\mathcal{O}_3$ and $\mathcal{O}_4$, we can decompose $G^{21}_{34}(x)$ in (\ref{G1234}) into conformal blocks $A^{21}_{34, \mathcal{O}}(x)$,
\begin{equation}
G^{21}_{34}(x)=\sum_{\mathcal{O}}C_{12\mathcal{O}}C^{\mathcal{O}}_{~~34}A^{21}_{34, \mathcal{O}}(x)\,.
\end{equation}
Similarly, we can also decompose $G^{41}_{32}(x)$ in (\ref{G1432}) into conformal blocks $A^{41}_{32, \mathcal{O}}(x)$ by imposing OPE between $\mathcal{O}_3$ and $\mathcal{O}_2$,
\begin{equation}
G^{41}_{32}(x)=\sum_{\mathcal{O}}C_{14\mathcal{O}}C^{\mathcal{O}}_{~~32}A^{41}_{32, \mathcal{O}}(x)\,.
\end{equation}
Then,the bootstrap equation for the warped conformal field theory resulting from the crossing symmetry (\ref{crosssym}) is given by,
\begin{equation}\label{bootstrap}
\sum_{\mathcal{O}}C_{12\mathcal{O}}C^{\mathcal{O}}_{~~34}A^{21}_{34, \mathcal{O}}(x)=\sum_{\mathcal{O}}C_{14\mathcal{O}}C^{\mathcal{O}}_{~~32}A^{41}_{32, \mathcal{O}}(1-x)\,.
\end{equation}
Once we have the closed form expressions of the conformal blocks we can solve the bootstrap equation (\ref{bootstrap}) to find all possible consistent warped conformal invariant theories.

The full warped conformal block is the Virasoro-Kac-Moody block. As was argued by~\cite{Fitzpatrick:2015zha}, it is more convenient to label the basis using spectral flow invariant generators (\ref{ONM}). For the block with weight $\Delta$ and charge $Q=-Q_1-Q_2$, using (\ref{inv3pt}), the warped conformal block factorizes as a product of a Virasoro block and a U(1) block,
\bea
A^{\Delta,Q}(x)=\mathcal{V}(c-1,\Delta_i^{inv},\Delta^{inv}, x)\mathcal{V}_{P}(k,Q_i,x)\,,
\eea
where $\mathcal{V}_{P}(k,Q_i,x)$ is the Kac-Moody block, and $\mathcal{V}(c-1,\Delta_i^{inv},\Delta^{inv}, x)$ is the Virasoro block with central charge $c\rightarrow c-1$, external operators $\Delta_i\rightarrow \Delta_i^{inv}$ and intermediate primary operator $\Delta\rightarrow\Delta^{inv}$.

Similar to the discussion for holographic CFTs~\cite{Fitzpatrick:2014vua, Fitzpatrick:2015zha}, the large $c$ limit of WCFT is also relevant for holography. At large $c$, the Virasoro block becomes SL(2,R) block \cite{Zamolodchikov:1985ie, Zamolodchikov:1987}.  Taking into account the total charge conservation $\sum_{i=1}^4Q_i=0$, the warped conformal block $A^{\Delta, Q}(x)$ at large $c$ is,
\bea
A^{\Delta, Q}(x)&=&\delta_{\sum_{i=1}^4Q_i, 0}\delta_{Q_1+Q_2, -Q}\,x^{\Delta^{inv}} \\
&&\times
_2F_1(\Delta^{inv}-\Delta^{inv}_1+\Delta^{inv}_2, \Delta^{inv}+\Delta^{inv}_3-\Delta^{inv}_4, 2\Delta^{inv}; x)\,\mathcal{V}_{P}(k,Q_i,x)\,.\nonumber
\eea

\section{Thermal correlators}\label{sec3}
Finite temperature results can be obtained from a warped conformal mapping,
\begin{equation}\label{ftm}
x=x'=e^{\frac{2\pi X}{\beta}},~~~~y=y'+qx'=Y+\bar{q}X\,.
\end{equation}
where $(x',\,y')$ are coordinates with spectral flow parameter $q$, $(X, Y)$ stands for the finite temperature coordinates, $\beta$ is the inverse temperature along $X$, and $\bar{q}$ is a constant which is related to the thermal identification along $Y$. The warped conformal generators (\ref{wg}) on the original plane  can be rewritten as (\ref{lpq}), and furthermore rewritten in the finite temperature coordinates,
\begin{equation}\label{wgft}
L_n=-\frac{\beta}{2\pi}e^{2\pi n\frac{X}{\beta}}(\partial_{X}-\bar{q}\partial_{Y}),~~~~P_n=ie^{2\pi n\frac{X}{\beta}}\partial_{Y}\,.
\end{equation}
At finite temperature, two point correlator for two identical primary operators $\mathcal{O}(X, Y)$ and $\mathcal{O}(0, 0)$ with conformal dimension $\Delta$ and charge $Q$ can be obtained from the transformation rule (\ref{ftG}) applying on the zero temperature result (\ref{2pf0}),
\begin{equation}\label{coord2pf}
G(X, Y)=\langle\mathcal{O}(X, Y)\mathcal{O}(0, 0)\rangle\sim(-1)^{\Delta}e^{iQ\left(Y+\bar{q}X\right)}\left(\frac{\beta}{\pi}\sinh\frac{\pi X}{\beta}\right)^{-2\Delta}\,,
\end{equation}
In momentum space, the retarded Green's functions are related to the two point functions (\ref{coord2pf}) by a Fourier transformation. In order to perform the Fourier transformation on (\ref{coord2pf}), we shall consider the ``Euclidean'' version of the correlator $G_E$ by Wick rotating the coordinate $X$, i.e., $X=iX_E$, namely, $G_E(X_E, Y)=G(iX_E, Y)$. $X_E$ has period $\beta$ and the momentum space Euclidean correlator is given by,
\begin{equation}
G_E(\omega_E)=\int_0^{\beta}\mathrm{d}X_Ee^{-i\omega_EX_E}G_E(X_E, Y)\,,
\end{equation}
where $\omega_E$ is the Euclidean momentum and it is related to the real momentum $\omega$ conjugates to $X$ through $\omega=-i\omega_E$. The above integral is divergent but can be defined by analytic continuation~\cite{Maldacena:1997ih},
\begin{equation}
G_E(\omega_E)\sim\beta^{1-2\Delta}e^{i\frac{\beta(\omega_E-iQ\bar{q})}{2}}\Gamma\left(\Delta+\frac{\beta(\omega_E-iQ\bar{q})}{2\pi}\right)\Gamma\left(\Delta-\frac{\beta(\omega_E-iQ\bar{q})}{2\pi}\right)\,.
\end{equation}
Note that, at finite temperature, $\omega_E$ takes discrete values of the Matsubara frequencies,
\begin{equation}
\omega_E=\frac{2\pi m}{\beta}\,,
\end{equation}
where $m$ is integer for bosonic modes and half integer for fermionic modes. We can find that for both integer and half integer $m$, the momentum dependence in the Euclidean correlator only comes from the Gamma functions,
\begin{equation}\label{GE}
G_E(\omega_E)\sim\beta^{1-2\Delta}\frac{e^{\frac{Q\bar{q}\beta}{2}}}{\sin\left(\pi\Delta-\frac{iQ\bar{q}\beta}{2}\right)}\frac{\Gamma\left(\Delta-\frac{\beta(\omega_E-iQ\bar{q})}{2\pi}\right)}{\Gamma\left(1-\Delta-\frac{\beta(\omega_E-iQ\bar{q})}{2\pi}\right)}\,.
\end{equation}
The retarded Green's function $G_R(\omega)$ is then obtained from analytic continuation from the Euclidean correlator $G_E(\omega_E)$, i.e.,
\begin{equation}\label{RE}
G_R(i\omega_E)=G_E(\omega_E),~~\omega_E>0\,.
\end{equation}
More explicitly, the retarded Green's function for the primary operator in WCFT takes the form,
\begin{equation}\label{GRb}
G_R(\omega)\sim\beta^{1-2\Delta}\frac{e^{\frac{Q\bar{q}\beta}{2}}}{\sin\left(\pi\Delta-\frac{iQ\bar{q}\beta}{2}\right)}\frac{\Gamma\left(\Delta+i\frac{\omega+Q\bar{q}}{2\pi/\beta}\right)}{\Gamma\left(1-\Delta+i\frac{\omega+Q\bar{q}}{2\pi/\beta}\right)}\,.
\end{equation}

\section{Twist Field and R$\acute{\textbf{e}}$nyi Entropy}\label{sec4}
In a CFT, the two point function of twist operators can be used to calculate entanglement entropy and
R$\acute{\mathrm{e}}$nyi entropy~\cite{Calabrese:2004eu, Calabrese:2009qy}.
This idea has been extended to GCFTs~\cite{Bagchi:2014iea, Basu:2015evh} and WCFTs~\cite{Castro:2015csg, Song:2016gtd}.
As a consistent check, here in this section we revisit the calculation of R$\acute{\mathrm{e}}$nyi entropy for WCFTs using the previously derived two point functions, and find that it is consistent with the result obtained in~\cite{Song:2016gtd}. Comparing to~\cite{Castro:2015csg}, there is an additional parameter $\alpha$. We will comment on this in more details around Eq. (\ref{cdc}).

More precisely, for two dimensional field theories with sufficiently constraining symmetries, such as CFTs, GCFTs, and WCFTs, the R$\acute{\mathrm{e}}$nyi entropy for an interval $\mathcal{D}$ can be written as,
%defined by taking the trace on $n$th power of the reduced density matrix $\rho_{\mathcal{D}}$,
\bea
S_n&=&\frac{1}{1-n}\log\frac{\mathrm{tr}(\rho_{\mathcal{D}}^n)}{(\mathrm{tr}\rho_{\mathcal{D}})^n}\,=\frac{1}{1-n}\log\frac{\langle\Phi_n(X_1)\Phi_n^{\dagger}(X_2)\rangle_{\mathcal{C}}}{\langle\Phi_1(X_1)\Phi_1^{\dagger}(X_2)\rangle_{\mathcal{C}}^n}\,.
\eea
Here in the first equality, the R$\acute{\mathrm{e}}$nyi entropy is related to the $n$th power of the reduced density matrix $\rho_{\mathcal{D}}$ for $\mathcal{D}$. This can be realized as a path integral an a manifold $\mathcal{R}_n$ which is made up of $n$ decoupled copies of the original space $\mathcal{R}_1$. In the second equality, $\Phi_n$ is the twist field inserted at the endpoints of the interval that enforce the replica boundary conditions on a plane $\mathcal{C}$. $X_{1, 2}$ are the endpoint coordinates of the interval $\mathcal{D}$.

With the assumption that the twist operators are primary operators, the two point function is fixed by the global symmetry, which depends on the conserved charges of $\Phi_n$. The charges can be further determined by noting that there are two different approaches for evaluating the expectation values of an operator on $\mathcal{R}_n$,
\begin{equation}\label{ExpO}
\langle\mathcal{O}(X^{(i)})\rangle_{\mathcal{R}_n}=\frac{\langle\mathcal{O}(X)\Phi_n(X_1)\Phi^{\dagger}_n(X_2)\rangle_{\mathcal{C}}}{\langle\Phi_n(X_1)\Phi^{\dagger}_n(X_2)\rangle_{\mathcal{C}}}\,,
\end{equation}
When $\mathcal O$ is a conserved current, the right hand side can be calculated by Ward identity,
and the left hand side can be calculated by the transformation rules or the Rindler method~\cite{Castro:2015csg}.

As to the WCFT, we have the $U(1)$ Kac-Moody current instead of a anti-holomorphic energy momentum tensor. The twist field thus is a charged field, and the Ward identities for the holomorphic energy momentum tensor $T(X)$ and the $U(1)$ current $P(X)$ help us to find out the explicit expressions for the conformal dimension and charge of the twist field $\Phi_n(X, Y)$. Let $Y$ denote the classically $U(1)$ preferred axis and $X$ the quantum anomaly selected axis with a scaling $SL(2,R)$ symmetry. The field theory in general has a thermal identification,
\begin{equation}\label{thermalid}
(X, Y)\sim(X+i\beta, Y-i\bar{\beta})\,.
\end{equation}
Consider an arbitrary inteval $\mathcal{D}$,
\begin{equation}\label{intv}
\mathcal{D}:~~~(X, Y)\in\left[\left(-\frac{\Delta X}{2}, -\frac{\Delta Y}{2}\right), \left(\frac{\Delta X}{2}, \frac{\Delta Y}{2}\right)\right]\,,
\end{equation}
and the Rindler transformation~\cite{Song:2016gtd},
\begin{equation}\label{Xtox}
\frac{\tanh\frac{\pi X}{\beta}}{\tanh\frac{\Delta X\pi}{2\beta}}=\tanh\frac{\pi\tilde{X}}{\kappa},~~~~Y+\left(\frac{\bar{\beta}}{\beta}-\frac{\alpha}{\beta}\right)X=\tilde{Y}+\left(\frac{\bar{\kappa}}{\kappa}-\frac{\alpha}{\kappa}\right)\tilde{X}\,.
\end{equation}
where $\alpha$, $\kappa$, and $\bar{\kappa}$ are arbitrary constants. The new coordinates $(\tilde{X}, \tilde{Y})$ covers a strip region $-{\Delta X\over2}<X<{\Delta X\over 2}$.
Under the Rindler transformation (\ref{Xtox}), the vectors $\partial_{\tilde{X}}$ and $\partial_{\tilde{Y}}$ should be linear combinations of the generators of global warped conformal symmetry at finite temperature~\cite{Jiang:2017ecm}. From (\ref{Xtox}), we find that,
\begin{eqnarray}
\partial_{\tilde{X}}&=&\frac{\bar{\kappa}-\alpha}{\kappa}\partial_Y-\frac{\beta}{\kappa}\left(\frac{\cosh\frac{2\pi X}{\beta}-\cosh\frac{\pi\Delta X}{\beta}}{\sinh\frac{\pi\Delta X}{\beta}}\right)\left(\partial_X-\frac{\bar{\beta}-\alpha}{\beta}\partial_Y\right),\nonumber\\
\\
\partial_{\tilde{Y}}&=&\partial_Y\,.
\end{eqnarray}
It is clear to find that
\begin{equation}\label{tq}
\bar{q}=\frac{\bar{\beta}-\alpha}{\beta}
\end{equation}
in order for $\partial_{\tilde{X}}$ and $\partial_{\tilde{Y}}$ to be in linear combinations of the global generators $\{L_{0, \pm1}, P_0\}$ given in (\ref{wgft}).

The transformation (\ref{Xtox}) is a warped conformal transformation, and hence is implemented by a unitary operator $U$.
The reduced density matrix $\rho_{\mathcal{D}}$ on $\mathcal D$ is given by $\rho_{\mathcal{D}}=U\rho_{\mathcal{H}}U^{\dagger}$, where $\rho_\mathcal{H}$ denotes the thermal density matrix after the transformation and $\mathcal{H}$ stands for the Rindler space. The expectation values of the energy momentum tensor $T(X)$ and $U(1)$ Kac-Moody current $P(X)$ can be calculated through the following formulas,
\begin{eqnarray}
\label{<T>}\langle T(X^{(i)})\rangle_{\mathcal{R}_n}&=&\frac{\mathrm{tr}(T(X)\rho_{\mathcal{D}}^n)}{\mathrm{tr}(\rho_{\mathcal{D}}^n)}=\frac{\mathrm{tr}(U^{\dagger}T(X)U\rho_{\mathcal{H}}^n)}{\mathrm{tr}(\rho_{\mathcal{H}}^n)}\,,\\
\label{<P>}\langle P(X^{(i)})\rangle_{\mathcal{R}_n}&=&\frac{\mathrm{tr}(P(X)\rho_{\mathcal{D}}^n)}{\mathrm{tr}(\rho_{\mathcal{D}}^n)}=\frac{\mathrm{tr}(U^{\dagger}P(X)U\rho_{\mathcal{H}}^n)}{\mathrm{tr}(\rho_{\mathcal{H}}^n)}\,.
\end{eqnarray}
The finite transformation properties  of $T$ (\ref{Tp}) and $P$ (\ref{Pp}) can be rewritten as
\begin{eqnarray}\label{UtranT}
U^{\dagger}T(X)U&=&\left(\frac{\partial\tilde{X}}{\partial X}\right)^2\left(T(\tilde{X})-\frac{c}{12}\{X, \tilde{X}\}\right)+\frac{\partial\tilde{X}}{\partial X}\frac{\partial\tilde{Y}}{\partial X}P(\tilde{X})-\frac{k}{4}\left(\frac{\partial\tilde{Y}}{\partial X}\right)^2\,,\\\label{UtranP}
U^{\dagger}P(X)U&=&\left(\frac{\partial\tilde{X}}{\partial X}\right)\left(P(\tilde{X})+\frac{k}{2}\frac{\partial Y}{\partial\tilde{X}}\right)\,.
\end{eqnarray}
The $n$th power of the thermal density matrix $\rho_{\mathcal{H}}$ is related to the partition function in $(\tilde{X}, \tilde{Y})$ coordinate system,
\begin{equation}
\mathrm{tr}(\rho_{\mathcal{H}}^n)=\mathrm{tr}_{\bar{a},a}\left(e^{2\pi in\bar{\theta}\mathcal{P}_0}e^{-2\pi in\theta \mathcal{L}_0}\right)=Z_{\bar{a}|a}(n\bar{\theta}|n\theta)\,,
\end{equation}
where $\mathcal{P}_0$ and $\mathcal{L}_0$ are the zero-modes of the energy momentum tensor and Kac-Moody current respectively, and the partition function $Z_{\bar{a}|a}(\bar{\theta}|\theta)$ is defined on a torus with spatial circle parameterized by $(\bar{a}, a)$ and thermal circle parameterized by $(\bar{\theta}, \theta)$,
\begin{equation}
(\tilde{X}, \tilde{Y})\sim(\tilde{X}+2\pi a, \tilde{Y}-2\pi\bar{a})\sim(\tilde{X}-2\pi\theta, \tilde{Y}+2\pi\bar{\theta})\,.
\end{equation}
In reference~\cite{Song:2016gtd}, the partition function in $(\tilde{X}, \tilde{Y})$ system has been calculated by using the coordinate transformation (\ref{Xtox}) and modular transformations,
\begin{equation}\label{partition}
Z_{\bar{a}|a}(\bar{\theta}|\theta)=\exp\left(\frac{\pi ik\bar{\theta}^2a}{2\theta}-2\pi i\left(\frac{a\bar{\theta}}{\theta}-\bar{a}\right)\mathcal{P}_0^{vac}+\frac{2\pi ia}{\theta}\mathcal{L}_0^{vac}\right)\,,
\end{equation}
where
\begin{equation}
2\pi a =\frac{\kappa\zeta}{\pi}, \quad 2\pi\bar{a}=\frac{\bar{\kappa}-\alpha}{\pi}\zeta-\Delta T-\Delta X(\frac{\bar{\beta}-\alpha}{\beta}), \quad 2\pi\theta=-i\kappa, \quad 2\pi\bar{\theta}=- i\bar{\kappa}\,,
\end{equation}
$k$ is the Kac-Moody level, $\mathcal{L}_0^{vac}$ and $\mathcal{P}_0^{vac}$ are the vacuum values of the zero-modes charges $\mathcal{L}_0$ and $\mathcal{P}_0$ respectively, and $\zeta$ is related to the UV cut-off $\epsilon$ in WCFT as,
\begin{equation}
\zeta=\log\left(\frac{\beta}{\pi\epsilon}\sinh\frac{\pi\Delta X}{\beta}\right)+\mathcal{O}(\epsilon)\,.
\end{equation}
In a translational invariant state, the values of the currents are simply related to the zero modes, i.e. $\mathcal{L}_0=-aT(\tilde{X})$ and $\mathcal{P}_0=-aP(\tilde{X})$. The expectation values of the currents in $(\tilde{X}, \tilde{Y})$ coordinate system can be written as,
\begin{equation}\label{Tx}
\langle T(\tilde{X}^{(i)})\rangle=\frac{1}{2\pi ian}\frac{\partial}{\partial\theta}\log Z_{\bar{a}|a}(n\bar{\theta}|n\theta),~~~~\langle P(\tilde{X}^{(i)})\rangle=-\frac{1}{2\pi ian}\frac{\partial}{\partial\bar{\theta}}\log Z_{\bar{a}|a}(n\bar{\theta}|n\theta)\,.
\end{equation}
Plugging the partition function (\ref{partition}) into the expectation values (\ref{Tx}), and the transformation rules (\ref{UtranT}) and (\ref{UtranP}) into (\ref{<T>}) and (\ref{<P>}), we get the expectation values of $T(X)$ and $P(X)$ on $\mathcal{R}_n$.

On the other hand, according to (\ref{ExpO}) the expectation values of $T(X)$ and $P(X)$ can also be obtained from the Ward identities,
\begin{equation}\label{widTP}
\langle T(X^{(i)})\rangle_{\mathcal{R}_n}=\frac{\langle T(X)\Phi_n(X_1, Y_1)\Phi_n^{\dagger}(X_2, Y_2)\rangle_{\mathcal{C}}}{\langle\Phi_n(X_1, Y_1)\Phi_n^{\dagger}(X_2, Y_2)\rangle_{\mathcal{C}}},~~~~\langle P(X^{(i)})\rangle_{\mathcal{R}_n}=\frac{\langle P(X)\Phi_n(X_1, Y_1)\Phi_n^{\dagger}(X_2, Y_2)\rangle_{\mathcal{C}}}{\langle\Phi_n(X_1, Y_1)\Phi_n^{\dagger}(X_2, Y_2)\rangle_{\mathcal{C}}}\,,
\end{equation}
Substituting the two point function for the twist operator $\Phi_n(X, Y)$ with dimension $\Delta_n$ and charge $Q_n$ at finite temperature,
\begin{equation}\label{2pfwithbq}
\langle\Phi_n(X_1, Y_1)\Phi_n^{\dagger}(X_2, Y_2)\rangle_{\mathcal{C}}\sim e^{iQ_n\left(Y_1-Y_2+\frac{\bar{\beta}-\alpha}{\beta}(X_1-X_2)\right)}\left(\frac{\beta}{\pi}\sinh\frac{\pi(X_1-X_2)}{\beta}\right)^{-2\Delta_n}\,.
\end{equation}
we get second expressions of the expectation values of $T(X)$ and $P(X)$ on $\mathcal{R}_n$. Comparing to the first expressions obtained from the Rindler transformation, we can find the explicit expressions for $\Delta_n$ and $Q_n$.

We will do the explicit calculations in the zero temperature case, by taking the zero temperature limit of (\ref{Xtox}), i.e.  $\bar{\beta}\to\infty$, $\beta\to\infty$, with $\bar{\beta}/\beta$ fixed.
This zero temperature limit will in general lead to a plane with spectral flow parameter $q={\bar{\beta}-\alpha\over\beta}$.  We should use the $(x',y')$ coordinates as in (\ref{lpq}) and (\ref{ftm}).
With some abuse of notation, in this subsection we will drop the prime, and use $(x,y)$  instead. According to the argument below Eq. (\ref{Tx}), the expectation values of $T(x)$ and $P(x)$ on $\mathcal{R}_n$ can now be written down after some algebras,
\begin{eqnarray}\label{0<TX>}
\langle T(x^{(i)})\rangle_{\mathcal{R}_n}&=&\frac{\Delta x^2}{\left(x-\frac{\Delta x}{2}\right)^2\left(x+\frac{\Delta x}{2}\right)^2}\left(\frac{c}{24}+\frac{\mathcal{L}_0^{vac}}{n^2}+\frac{i\mathcal{P}_0^{vac}\alpha}{2n\pi}-\frac{\alpha^2k}{16\pi^2}\right)\nonumber\\
&&+\frac{\Delta x}{\left(x-\frac{\Delta x}{2}\right)\left(x+\frac{\Delta x}{2}\right)}\frac{\bar{\beta}}{\beta}\left(-\frac{i\mathcal{P}_0^{vac}}{n}+\frac{k\alpha}{4\pi}\right)-\frac{k}{4}\frac{\bar{\beta}^2}{\beta^2}\,,\\\label{0<PX>}
\langle P(x^{(i)})\rangle_{\mathcal{R}_n}&=&\frac{\Delta x}{\left(x-\frac{\Delta x}{2}\right)\left(x+\frac{\Delta x}{2}\right)}\left(-\frac{i\mathcal{P}_0^{vac}}{n}+\frac{k\alpha}{4\pi}\right)-\frac{k}{2}\frac{\bar{\beta}}{\beta}\,.
\end{eqnarray}
On the other hand, by using the Ward identities for the energy momentum tensor $T(x)$ and the $U(1)$ current $P(x)$~\cite{cwi}
\begin{eqnarray}
&&\langle T(x)\Phi_n(x_1, y_1)\Phi_n^{\dagger}(x_2, y_2)\rangle_{\mathcal{C}}\nonumber\\
&=&\sum_{i=1}^2\left(\frac{\Delta_n/n}{(x-x_i)^2}+\frac{1/n}{x-x_i}\frac{\partial}{\partial x_i}+\langle T(x)\rangle_{\mathcal{C}}\right)\langle\Phi_n(x_1, y_1)\Phi_n^{\dagger}(x_2, y_2)\rangle_{\mathcal{C}}\,,
 \\ \nonumber\\
&&
\nonumber \langle P(x)\Phi_n(x_1, y_1)\Phi_n^{\dagger}(x_2, y_2)\rangle_{\mathcal{C}}\\
&=&\sum_{i=1}^{2}\left(\frac{1/n}{x-x_i}\frac{\partial}{\partial y_i}+\langle P(x)\rangle_{\mathcal{C}}\right)\langle\Phi_n(x_1, y_1)\Phi_n^{\dagger}(x_2, y_2)\rangle_{\mathcal{C}}
\end{eqnarray}
in (\ref{widTP}) and substituting the two point function (\ref{2pfwithbq}) in the zero temperature limit, we find that,
\begin{eqnarray}\label{0<TX>tw}
\langle T(x^{(i)})\rangle_{\mathcal{R}_n}&=&\frac{\Delta x^2}{\left(x-\frac{\Delta x}{2}\right)^2\left(x+\frac{\Delta x}{2}\right)^2}\frac{\Delta_n}{n}+\frac{\Delta x}{\left(x-\frac{\Delta x}{2}\right)\left(x+\frac{\Delta x}{2}\right)}\frac{\bar{\beta}}{\beta}\frac{iQ_n}{n}+\langle T(x)\rangle_{\mathcal{C}}\,,\\\label{0<PX>tw}
\langle P(x^{(1)})\rangle_{\mathcal{R}_n}&=&\frac{\Delta x}{\left(x-\frac{\Delta x}{2}\right)\left(x+\frac{\Delta x}{2}\right)}\frac{iQ_n}{n}+\langle P(x)\rangle_{\mathcal{C}}\,.
\end{eqnarray}
Comparing Eqs. (\ref{0<TX>tw}) and (\ref{0<PX>tw}) to that Eqs. (\ref{0<TX>}) and (\ref{0<PX>}), we get the expressions for the conformal dimension and charge of the twist field,
\begin{equation}\label{cdc}
\triangle_n=n\left(\frac{c}{24}+\frac{\mathcal{L}_0^{vac}}{n^2}+\frac{i\mathcal{P}_0^{vac}\alpha}{2n\pi}-\frac{\alpha^2k}{16\pi^2}\right),~~~~Q_n=n\left(-\frac{\mathcal{P}_0^{vac}}{n}-i\frac{k\alpha}{4\pi}\right)\,.
\end{equation}
In contrast to the result in~\cite{Castro:2015csg}, there is an additional $\alpha$ parameter. The value of $\alpha$ can not be determined in field theory side without knowing more information of the WCFT. As stated in~\cite{Song:2016gtd}, $\alpha$ is proportional to the slop of the thermal identification after modular transformation. Comparing to results in the literature, taking finite $\alpha$ is in fact the slow rotating limit described in section 3.1 of~\cite{Detournay:2012pc}. In addition, we find that
\begin{equation}
\langle T(x)\rangle_{\mathcal{C}}=-\frac{k}{4}\frac{\bar{\beta}^2}{\beta^2},~~~~\langle P(x)\rangle_{\mathcal{C}}=-\frac{k}{2}\frac{\bar{\beta}}{\beta}\,.
\end{equation}
which is just the charges on the plane with a spectral flow parameter $q={\bar\beta-\alpha\over\beta}$ at this limit.
Once we write the conformal dimension and charge of the twist field in terms of the central charge $c$, level $k$, and vacuum values of the zero-modes charges $L_0$ and $P_0$, it is straightforward to find out the R$\acute{\mathrm{e}}$nyi entropy as a function of such constants and the interval (\ref{intv}),
\begin{eqnarray}
S_n&=&\frac{1}{1-n}\log\frac{\mathrm{tr}(\rho_{\mathcal{D}}^n)}{(\mathrm{tr}\rho_{\mathcal{D}})^n}=\frac{1}{1-n}\log\frac{\langle\Phi_n(x_1, y_1)\Phi_n^{\dagger}(x_2, y_2)\rangle_{\mathcal{C}}}{\langle\Phi_1(x_1, y_1)\Phi_1^{\dagger}(x_2, y_2)\rangle_{\mathcal{C}}^n}\nonumber\\
&=&-i\mathcal{P}_0^{vac}\left(\Delta y+\frac{\bar{\beta}}{\beta}\Delta x\right)+\left(-\frac{\alpha}{\pi}i\mathcal{P}_0^{vac}-\frac{2(n+1)}{n}\mathcal{L}_0^{vac}\right)\log\Delta x\,.
\end{eqnarray}
This result matches the zero temperature limit of the R$\acute{\mathrm{e}}$nyi entropy calculated in~\cite{Song:2016gtd}.
For finite temperature, it is easy to check that this method also reproduce the result of~\cite{Song:2016gtd}.

\section{Retarded Green's Functions in WAdS/WCFT}\label{sec5}
%Based on the AdS/CFT correspondence~\cite{Maldacena:1997re, Gubser:1998bc, Witten:1998qj}, a finite temperature conformal field theory is supposed to be dual to a bulk asymptotically AdS geometry that involves black holes. Under this duality, a CFT scalar operator on the boundary is dual to a bulk scalar field with mass relating to the conformal dimension of the scalar operator. The two point functions for the boundary operators is dual to the retarded Green's functions in the bulk \cite{Liu}. The \wei{retarded Green's functions }{ is it bulk or boundary} are related to the field configurations with ingoing boundary conditions near the horizons of the black holes, while advanced Green's functions are related to outgoing boundary conditions.

From a historical perspective, the study of WCFT was, at least partially, inspired by the study of holography for WAdS and Kerr spacetimes. With Dirichlet-Neumann boundary conditions~\cite{Compere:2009zj}, WAdS is conjectured to be holographically dual to a WCFT \cite{Detournay:2012pc}. With Dirichlet boundary conditions~\cite{Compere:2014bia}, WAdS is conjectured~\cite{Anninos:2008fx} to be holographically dual to a CFT. Evidence for WAdS/CFT include the microscopic interpretation of black hole entropy~\cite{Anninos:2008fx}, two point correlation functions~\cite{Bredberg:2009pv, Chen:2009cg}, and holographic entanglement entropy~\cite{Song:2016pwx}. Evidence for WAdS/WCFT include the microscopic interpretation of black hole entropy \cite{Detournay:2012pc}, and holographic entanglement entropy~\cite{Song:2016gtd}. In this section, we will revisit the scattering problem in WAdS, and give a prescription for the retarded Green's function in the WAdS/WCFT correspondence. Similar prescription can be also applied to Kerr spacetime, suggesting the possibility of Kerr has a holographic dual as WCFT.

Note that from the WCFT analysis, $\partial_Y$ is the generator of $U(1)$. This suggest that in WAdS, momentum along the $Y$ direction should be viewed as a charge, and kept fixed. This is the main difference from the WAdS/CFT story, where this momentum is viewed as Fourier mode.
Viewing momentum as a charge also provides a solution to the puzzle that conformal dimension depends on momentum first encountered in~\cite{Bredberg:2009pv}.

The warped AdS geometry that we are considering is the black string solution of the S-dual dipole truncation from type IIB supergravity~\cite{Detournay:2012dz}. The finite temperature warped black string metric can be written in a coordinate system $(X, Y, \rho)$ that is directly related to the boundary WCFT coordinates $(X, Y)$ as~\cite{Song:2016gtd},
\begin{eqnarray}\label{ds2TUTV}
ds^2&=&\ell^2\Big((T_U^2(1+\Lambda^2T_V^2)-\Lambda^2\rho^2)dX^2+2\rho dX\left(\frac{dY}{T_V}+dX\right)+T_V^2\left(\frac{dY}{T_V}+dX\right)^2\nonumber\\
&&+\frac{(1+\Lambda^2T_V^2)d\rho^2}{4(\rho^2-T_U^2T_V^2)}\Big)\,,
\end{eqnarray}
where $\Lambda$ stands for the warping parameter, $\ell$ is the AdS radius when $\Lambda=0$, and $T_U$ and $T_V$ are parameters that related to thermal identification. Under this background, we shall consider a massive scalar perturbation $\Phi$ with mass $m$ that satisfies the following equation of motion,
\begin{equation}\label{scalareomTX}
(\nabla_{\mu}\nabla^{\mu}-m^2)\Phi=0\,.
\end{equation}
A general solution with fixed momentum along $Y$ can be written as a Fourier transformation along $X$ direction,
\begin{equation}\label{ftq}
\Phi_{Q}=\int\mathrm{d}\omega~e^{i(QY-\omega X)}\phi_Q(\omega,\rho)\,,
\end{equation}
where $Q$ is held fixed. With this ansatz, the equation of motion of the scalar perturbation $\Phi$ (\ref{scalareomTX}) reduces to a radial equation for the Fourier mode $\phi_Q(P, z)$,
\begin{equation}\label{EOM}
\frac{\mathrm{d}}{\mathrm{d}z}\left((z^2-\frac{1}{4})\frac{\mathrm{d}}{\mathrm{d}z}\phi_Q(\omega, z)\right)+\left(\frac{k_+^2}{4\left(z-\frac{1}{2}\right)}-\frac{k_-^2}{4\left(z+\frac{1}{2}\right)}+\Delta_+\Delta_-\right)\phi_Q(\omega, z)=0\,,
\end{equation}
where $z=\rho/(2T_UT_V)$, and
\begin{eqnarray}
k_+&=&\frac{-\omega-Q\left(T_V+T_U\right)}{2T_U}\,,\label{k+}\\
k_-&=&\frac{-\omega-Q\left(T_V-T_U\right)}{2T_U}\,,\label{k-}\\
\Delta_{\pm}&=&\frac{1}{2}\pm\frac{1}{2}\sqrt{1+(1+\Lambda^2T_V^2)\ell^2m^2+\Lambda^2T_V^2Q^2}\,.\label{n}
\end{eqnarray}
The solution of (\ref{EOM}) with only ingoing modes near horizon $z=1/2$ can be written in the form of hypergeometric function,
\begin{eqnarray}
\phi_Q(\omega, z)&=&\left(z+\frac{1}{2}\right)^{-\Delta_+}\left(\frac{z-\frac{1}{2}}{z+\frac{1}{2}}\right)^{-\frac{ik_+}{2}}\nonumber\\
&\times&F\left(\Delta_+-i\left(\frac{k_+}{2}+\frac{k_-}{2}\right), \Delta_+-i\left(\frac{k_+}{2}-\frac{k_-}{2}\right), 1-ik_+; \frac{z-\frac{1}{2}}{z+\frac{1}{2}}\right)\,.
\end{eqnarray}
The asymptotic behavior of the above solution takes the form,
%\begin{eqnarray}
%\phi_Q(\omega, z)|_{z\to\infty}&\sim&z^{-\Delta_-}\frac{\pi\csc(2\pi\Delta_-)\Gamma(1+ik_+)}{\Gamma(2\Delta_-)\Gamma\left(\Delta_++\left(\frac{ik_+}{2}+\frac{ik_-}{2}\right)\right)\Gamma\left(\Delta_++\left(\frac{ik_+}{2}-\frac{ik_-}{2}\right)\right)}\nonumber\\
%&+&z^{-\Delta_+}\frac{\pi\csc(2\pi\Delta_+)\Gamma(1+ik_+)}{\Gamma(2\Delta_+)\Gamma\left(\Delta_-+\left(\frac{ik_+}{2}+\frac{ik_-}{2}\right)\right)\Gamma\left(\Delta_-+\left(\frac{ik_+}{2}-\frac{ik_-}{2}\right)\right)}\,.
%\end{eqnarray}
%By using of Eqs. (\ref{k+}) and (\ref{k-}), the above asymptotic expansion of $\phi_Q(\omega, z)$ can be written as,
\begin{equation}\label{asyexp}
\phi_Q(\omega, z)|_{z\to\infty}\sim\frac{\phi_Q^b(\omega)}{z^{\Delta_-}}-\frac{\phi_Q^b(\omega)}{z^{\Delta_+}}\frac{\Gamma(2\Delta_-)\Gamma\left(\Delta_++\frac{iQ}{2}\right)}{\Gamma(2\Delta_+)\Gamma\left(\Delta_-+\frac{iQ}{2}\right)}\frac{\Gamma\left(\Delta_++\frac{i(\omega+QT_V)}{2T_U}\right)}{\Gamma\left(\Delta_-+\frac{i(\omega+QT_V)}{2T_U}\right)}\,,
\end{equation}
where $\phi_Q^b(\omega)$ stands for the boundary field in momentum space. Then, the retarded Green's function for the scalar perturbation $\Phi$ in momentum space can be read of from the asymptotic form (\ref{asyexp})~\cite{Iqbal:2009fd, Chen:2009cg, Chen:2010ni},
\begin{equation}\label{Grbk}
G_R^{bk}\sim-\frac{\Gamma(2\Delta_-)\Gamma\left(\Delta_++\frac{iQ}{2}\right)}{\Gamma(2\Delta_+)\Gamma\left(\Delta_-+\frac{iQ}{2}\right)}\frac{\Gamma\left(\Delta_++\frac{i(\omega+QT_V)}{2T_U}\right)}{\Gamma\left(\Delta_-+\frac{i(\omega+QT_V)}{2T_U}\right)}\,,
\end{equation}
where $\sim$ stands for equaling up to a factor independent of $\omega$ and $Q$. (\ref{Grbk}) is the same momentum result as in~\cite{Bredberg:2009pv, Chen:2009cg}. To get to position space, if both $\omega$ and $Q$ are integrated, (\ref{Grbk}) leads to a retarded Green's function of a two dimensional CFT~\cite{Bredberg:2009pv, Chen:2009cg}. Here in WCFT, $Q$ is fixed in (\ref{ftq}), and we only need to integrate $\omega$ when going back to position space. We can absorb all $\omega$ independent factors into the normalization. In this sense, the retarded Green's function (\ref{Grbk}) can be rewritten as,
\begin{equation}\label{Grrbk}
G_R^{bk}\sim\frac{\Gamma\left(\Delta_++\frac{i(\omega+QT_V)}{2T_U}\right)}{\Gamma\left(1-\Delta_++\frac{i(\omega+QT_V)}{2T_U}\right)}\,,
\end{equation}
The above bulk result (\ref{Grrbk}) matches the boundary retarded Green's function (\ref{GRb}) with the dictionary~\cite{Song:2016gtd},
\begin{equation}
\Delta_+=\Delta,\quad T_U=\frac{\pi}{\beta},~~~~T_V=\bar{q}\,.
\end{equation}

\appendix
\section{Primary Representations}\label{appA}
In this appendix, we rederive (\ref{Pn} ) and (\ref{P0Q}) by a different approach. We consider primary representation of the warped conformal algebra at the origin.
A primary operator at $(0,0)$, with conformal dimension $\Delta$ and charge $Q$  can be defined as,
\bea
\,[\mathcal{L}_n, \mathcal{O}(0, 0)]&=&[\mathcal{P}_n, \mathcal{O}(0, 0)]=0, \quad n>0\\
\,[\mathcal{L}_0, \mathcal{O}(0, 0)]&=&\Delta\mathcal{O}(0, 0),~~~~[\mathcal{P}_0, \mathcal{O}(0, 0)]=-Q\mathcal{O}(0, 0)\,.
\eea
Local operators at arbitrary points can be introduced by unitary transformations, similar to ~\cite{Bagchi:2009ca, Bagchi:2010vw},
\begin{equation}\label{Oxt}
\mathcal{O}(x, y)=U^{-1}\mathcal{O}(0, 0)U,~~~~\text{where}~~U=e^{-x\mathcal{L}_{-1}-iy\mathcal{P}_0}\,.
\end{equation}
Now let us calculate the commutators $[\mathcal{L}_n, \mathcal{O}(x, y)]$ and $[\mathcal{P}_n, \mathcal{O}(x, y)]$ for any $n\geq0$,
\begin{align}\label{LnO}
[\mathcal{L}_n, \mathcal{O}(x, y)]=&[\mathcal{L}_n, U^{-1}\mathcal{O}(0, 0)U]\nonumber\\
=&[\mathcal{L}_n, U^{-1}]\mathcal{O}(0, 0)U+U^{-1}\mathcal{O}(0, 0)[\mathcal{L}_n, U]+U^{-1}[\mathcal{L}_n, \mathcal{O}(0, 0)]U\nonumber\\
=&U^{-1}[U\mathcal{L}_nU^{-1}-\mathcal{L}_n, \mathcal{O}(0, 0)]U+\delta_{n, 0}\Delta\mathcal{O}(x, y)\,,\\
\label{PnO}
[\mathcal{P}_n, \mathcal{O}(x, y)]=&[\mathcal{P}_n, U^{-1}\mathcal{O}(0, 0)U]\nonumber\\
=&[\mathcal{P}_n, U^{-1}]\mathcal{O}(0, 0)U+U^{-1}\mathcal{O}(0, 0)[\mathcal{P}_n, U]+U^{-1}[\mathcal{P}_n, \mathcal{O}(0, 0)]U\nonumber\\
=&U^{-1}[U\mathcal{P}_nU^{-1}-\mathcal{P}_n, \mathcal{O}(0, 0)]U-\delta_{n, 0}Q\mathcal{O}(x, y)\,.
\end{align}
Using the Baker-Campbell-Hausdorff formula and the warped conformal algebra (\ref{wcack}), one can show that,
\begin{eqnarray}
U\mathcal{L}_nU^{-1}&=&\sum_{k=0}^{n+1}\frac{(n+1)!}{(n+1-k)!k!}x^k\mathcal{L}_{n-k}\,,\\
U\mathcal{P}_nU^{-1}&=&\sum_{k=0}^{n}\frac{n!}{(n-k)!k!}x^k\mathcal{P}_{n-k}\,.
\end{eqnarray}
According to the definition of the local operators (\ref{Oxt}), one can show that,
\begin{equation}\label{-10}
[\mathcal{L}_{-1}, \mathcal{O}(x, y)]=\partial_x\mathcal{O}(x, y),~~~~[\mathcal{P}_0, \mathcal{O}(x, y)]=i\partial_y\mathcal{O}(x, y)\,.
\end{equation}
By using the above formulas and Eq. (\ref{prim}), we have
\begin{align}
[\mathcal{L}_n, \mathcal{O}(x, y)]=&[x^{n+1}\partial_x+(n+1)x^n\Delta]\mathcal{O}(x, y)\,,\\
[\mathcal{P}_n, \mathcal{O}(x, y)]=&-x^nQ\mathcal{O}(x, y)\,.
\end{align}
Here we have used the fact that U commutes with $\mathcal{L}_{-1}$ and $\mathcal{P}_0$ and any function of $x$ and $y$.

\section*{Acknowledgement}
We thank Alejandra Castro, Pankaj Chaturvedi, Monica Guica, Hongliang Jiang, Ning Su, Qiang Wen, and Jie-Qiang Wu for helpful discussions. This work was supported in part by start-up funding from Tsinghua University. The work of W.S. was also partially supported by the National Thousand-Young-Talents Program of China and a grant from the Simons Foundation. This work was completed while W.S. was participating workshops at the Aspen Center for Physics, which is supported by National Science Foundation grant PHY-1066293.

\end{document}